\documentclass[12pt,twoside]{article}
\usepackage{latexsym}

\iftrue
\def\,{\mskip 3mu} \def\>{\mskip 4mu plus 2mu minus 4mu} \def\;{\mskip 5mu plus 5mu} \def\!{\mskip-3mu}
\def\dispmuskip{\thinmuskip= 3mu plus 0mu minus 2mu \medmuskip=  4mu plus 2mu minus 2mu \thickmuskip=5mu plus 5mu minus 2mu}
\def\textmuskip{\thinmuskip= 0mu                    \medmuskip=  1mu plus 1mu minus 1mu \thickmuskip=2mu plus 3mu minus 1mu}
\textmuskip
\def\eqsp{\vspace{0ex}}
\def\beq{\dispmuskip\eqsp\begin{equation}}    \def\eeq{\eqsp\end{equation}\textmuskip}
\def\beqn{\dispmuskip\eqsp\begin{displaymath}}\def\eeqn{\eqsp\end{displaymath}\textmuskip}
\def\bqa{\dispmuskip\eqsp\begin{eqnarray}}    \def\eqa{\eqsp\end{eqnarray}\textmuskip}
\def\bqan{\dispmuskip\eqsp\begin{eqnarray*}}  \def\eqan{\eqsp\end{eqnarray*}\textmuskip}
\else
\def\beq{\begin{equation}}    \def\eeq{\end{equation}}
\def\beqn{\begin{displaymath}}\def\eeqn{\end{displaymath}}
\def\bqa{\begin{eqnarray}}    \def\eqa{\end{eqnarray}}
\def\bqan{\begin{eqnarray*}}  \def\eqan{\end{eqnarray*}}
\fi

\newtheorem{theorem}{Theorem}
\newtheorem{corollary}[theorem]{Corollary}
\newtheorem{lemma}[theorem]{Lemma}
\newtheorem{definition}[theorem]{Definition}
\newtheorem{proposition}[theorem]{Proposition}
\newtheorem{tablex}[theorem]{Table}
\newtheorem{figurex}[equation]{Figure}

\newenvironment{keywords}{\centerline{\bf\small
Keywords}\vspace{-1ex}\begin{quote}\small}{\par\end{quote}\vskip
1ex}

\def\ftheorem#1#2#3{\begin{theorem}[#2]\label{#1} #3 \end{theorem} }

\def\flemma#1#2#3{\begin{lemma}[#2]\label{#1} #3 \end{lemma} }
\def\fdefinition#1#2#3{\begin{definition}[#2]\label{#1} #3 \end{definition} }
\def\fproposition#1#2#3{\begin{proposition}[#2]\label{#1} #3 \end{proposition} }

\def\myparskip{\vspace{1.5ex plus 0.5ex minus 0.5ex}\noindent}
\def\paradot#1{\myparskip{\bfseries\boldmath{#1.}}}

\def\toinfty#1{\stackrel{#1\to\infty}{\longrightarrow}}
\def\nq{\hspace{-1em}}
\def\qed{\hspace*{\fill}$\Box\quad$\vspace{1ex plus 0.5ex minus 0.5ex}}
\def\odt{{\textstyle{1\over 2}}}
\def\odf{{\textstyle{1\over 4}}}
\def\eps{\varepsilon}                   
\def\epstr{\epsilon}                    
\def\qmbox#1{{\quad\mbox{#1}\quad}}

\def\leqa{\smash{\stackrel{\raisebox{1ex}{$\scriptstyle\!\!\;+$}}{\smash\leq}}}

\def\eqm{\smash{\stackrel{\raisebox{0.6ex}{$\scriptstyle\times$}}{\smash=}}}
\def\leqm{\smash{\stackrel{\raisebox{1ex}{$\scriptstyle\!\!\;\times$}}{\smash\leq}}}
\def\geqm{\smash{\stackrel{\raisebox{1ex}{$\scriptstyle\times\!\!\;$}}{\smash\geq}}}

\def\v{}
\def\l{{\ell}}                          
\def\M{{\cal M}}                        
\def\X{{\cal X}}                        
\def\E{{\bf E}}                         
\def\P{{\bf P}}                         
\def\B{\{0,1\}}                        

\def\MM{M}                              
\def\e{{\rm e}}                        
\def\a{\alpha}
\def\o{\omega}
\def\SetN{I\!\!N} \def\SetQ{I\!\!\!Q} \def\SetR{I\!\!R} 
\def\lb{\log}
\def\text#1{\mbox{\scriptsize{#1}}}

\begin{document}

\title{\vskip -25mm
\vskip 2mm\bf\LARGE\hrule height5pt \vskip 3mm
\sc On Semimeasures Predicting Martin-L\"of Random Sequences%
\thanks{This work was partially supported by the Swiss National Science Foundation
(SNF grants 2100-67712 and 200020-107616) and the Russian Foundation
for Basic Research (RFBR grants N04-01-00427 and N02-01-22001).
Submitted 15.May'05. Revised 26.Sep'06.
A shorter version appeared in the proceedings
of the ALT 2004 conference \cite{Hutter:04mlconvx}.}
\vskip 4mm \hrule height2pt \vskip 0mm}
\author{
{\bf Marcus Hutter}\\[3mm]
\normalsize IDSIA, Galleria 2, CH-6928\ Manno-Lugano, Switzerland\\
\normalsize RSISE/ANU/NICTA, \ Canberra, ACT, 0200, \ Australia\\
\normalsize marcus@hutter1.net \hspace{11ex} http://www.hutter1.net\\[5mm]
{\bf Andrej Muchnik}\\[3mm]
\normalsize Institute of New Technologies, 10 Nizhnyaya Radischewskaya\\
\normalsize Moscow 109004, Russia \hfill
\normalsize muchnik@lpcs.math.msu.su 
}
\date{14 October 2006}
\maketitle

\vspace{-5mm}
\begin{abstract}
\noindent
Solomonoff's central result on induction is that the prediction of
a universal semimeasure $\MM$ converges rapidly and with
probability 1 to the true sequence generating predictor $\mu$, if
the latter is computable. Hence, $M$ is eligible as a universal
sequence predictor in case of unknown $\mu$.
Despite some nearby results and proofs in the literature, the
stronger result of convergence for all (Martin-L{\"o}f) random
sequences remained open. Such a convergence result would be
particularly interesting and natural, since randomness can be
defined in terms of $\MM$ itself.
We show that there are universal semimeasures $M$ which do not
converge to $\mu$ on all $\mu$-random sequences, i.e.\ we give a
partial negative answer to the open problem.
We also provide a positive answer for some non-universal
semimeasures. We define the incomputable measure $D$ as a
mixture over all computable measures and the enumerable
semimeasure $W$ as a mixture over all enumerable nearly-measures.
We show that $W$ converges to $D$ and $D$ to $\mu$ on all random
sequences.
The Hellinger distance measuring closeness of two distributions
plays a central role.
\end{abstract}


\pagebreak
\section{Introduction}\label{secIntro}

\begin{quote}\it
``All difficult conjectures should be proved by reductio ad absurdum
arguments. For if the proof is long and complicated enough you are
bound to make a mistake somewhere and hence a contradiction will
inevitably appear, and so the truth of the original conjecture is
established QED.'' \par
\hfill --- {\sl Barrow's second `law' (2004)}
\end{quote}

A sequence prediction task is defined as to predict the next
symbol $x_n$ from an observed sequence $x=x_1...x_{n-1}$.
The key concept to attack general prediction problems is
Occam's razor, and to a less extent Epicurus' principle of
multiple explanations. The former/latter may be interpreted as to
keep the simplest/all theories consistent with the observations
$x_1...x_{n-1}$ and to use these theories to predict $x_n$.
Solomonoff \cite{Solomonoff:64,Solomonoff:78} formalized and
combined both principles in his universal a priori semimeasure $M$
which assigns high/low probability to simple/complex environments
$x$, hence implementing Occam and Epicurus. Formally it can be
represented as a mixture of all enumerable semimeasures.
An abstract characterization of $M$ by Levin \cite{Zvonkin:70} is
that $M$ is a universal enumerable semimeasure in the sense that
it multiplicatively dominates all enumerable semimeasures.

Solomonoff's \cite{Solomonoff:78} central result is that if the
probability $\mu(x_n|x_1...x_{n-1})$ of observing $x_n$ at time
$n$, given past observations $x_1...x_{n-1}$ is a computable
function, then the universal predictor $M_n:=M(x_n|x_1...x_{n-1})$
converges (rapidly!) {\em with $\mu$-probability 1} (w.p.1) for
$n\to\infty$ to the optimal/true/informed predictor
$\mu_n:=\mu(x_n|x_1...x_{n-1})$, hence $M$ represents a universal
predictor in case of unknown ``true'' distribution $\mu$.
Convergence of $M_n$ to $\mu_n$ w.p.1 tells us that $M_n$ is close
to $\mu_n$ for sufficiently large $n$ for almost all sequences
$x_1x_2...$. It says nothing about whether convergence is true for
any {\em particular} sequence (of measure 0).

Martin-L{\"o}f (M.L.) randomness is the standard notion for
randomness of individual sequences \cite{MartinLoef:66,Li:97}. A
M.L.-random sequence passes {\em all} thinkable effective randomness
tests, e.g.\ the law of large numbers, the law of the iterated
logarithm, etc. In particular, the set of all $\mu$-random sequences
has $\mu$-measure 1.
It is natural to ask whether $M_n$ converges to $\mu_n$ (in
difference or ratio) individually for all M.L.-random sequences.
Clearly, Solomonoff's result shows that convergence may at most
fail for a set of sequences with $\mu$-measure zero. A convergence
result for M.L.-random sequences would be particularly interesting and
natural in this context, since M.L.-randomness can be defined in
terms of $M$ itself \cite{Levin:73random}.
Despite several attempts to solve this problem
\cite{Vovk:87,Vitanyi:00,Hutter:03unipriors}, it remained open
\cite{Hutter:03mlconv}.

In this paper we construct an M.L.-random sequence and show the
existence of a universal semimeasure which does not converge on
this sequence, hence answering the open question negatively for
some $M$.
It remains open whether there exist (other) universal
semimeasures, probably with particularly interesting additional
structure and properties, for which M.L.-convergence holds.
The main positive contribution of this work is the construction of
a non-universal enumerable semimeasure $W$ which M.L.-converges to
$\mu$ as desired. As an intermediate step we consider the
incomputable measure $\hat D$, defined as a mixture over all
computable measures. We show M.L.-convergence of predictor $W$ to
$\hat D$ and of $\hat D$ to $\mu$.
The Hellinger distance measuring closeness of two predictive
distributions plays a central role in this work.

The paper is organized as follows:
In Section~\ref{secUniM} we give basic notation and results (for
strings, numbers, sets, functions, asymptotics, computability
concepts, prefix Kolmogorov complexity), and define and discuss
the concepts of (universal) (enumerable) (semi)measures.
Section~\ref{secConv} summarizes Solomonoff's and G\'acs' results
on predictive convergence of $M$ to $\mu$ with probability 1. Both
results can be derived from a bound on the expected Hellinger sum.
We present an improved bound on the expected exponentiated
Hellinger sum, which implies very strong assertions on the
convergence rate.
In Section~\ref{secMLNconv} we investigate whether convergence for
all Martin-L{\"o}f random sequences hold.
We construct a $\mu$-M.L.-random sequence on which some universal
semimeasures $M$ do not converge to $\mu$. We give a non-constructive
and a constructive proof of different virtue.
In Section~\ref{secMLconv} we present our main positive result. We
derive a finite bound on the Hellinger sum between $\mu$ and $\hat
D$, which is exponential in the randomness deficiency of the
sequence and double exponential in the complexity of $\mu$. This
implies that the predictor $\hat D$ M.L.-converges to $\mu$.
Finally, in Section~\ref{secNUESW} we show that $W$ is non-universal
and asymptotically M.L.-converges to $\hat D$, and summarize the
computability, measure, and dominance properties of $M$, $D$, $\hat
D$, and $W$.
Section~\ref{secDisc} contains discussion and outlook.

\section{Notation \& Universal Semimeasures $\MM$}\label{secUniM}

\noindent {\bf Strings.}
Let $i,k,n,t\in\SetN=\{1,2,3,...\}$ be natural numbers,
$x,y,z\in\X^*=\bigcup_{n=0}^\infty\X^n$ be finite strings of
symbols over finite alphabet $\X\ni a,b$.
We write $xy$ for the concatenation of string $x$ with $y$.
We denote strings $x$ of length $\l(x)=n$ by
$x=x_1x_2...x_n\in\X^n$ with $x_t\in\X$ and further abbreviate
$x_{k:n}:=x_k x_{k+1}...x_{n-1}x_n$ for $k\leq n$, and $x_{<n}:=x_1...
x_{n-1}$, and $\epstr=x_{<1}=x_{n+1:n}\in\X^0=\{\epstr\}$ for the
empty string. Let $\omega=x_{1:\infty}\in\X^\infty$ be a generic
and $\alpha\in\X^\infty$ a specific infinite sequence. For a
given sequence $x_{1:\infty}$ we say that $x_t$ is on-sequence and
$\bar x_t\neq x_t$ is off-sequence. $x'_t$ may be on- or
off-sequence.
We identify strings with natural numbers (including zero, $\X^*\cong
\SetN\cup\{0\}$).

\noindent {\bf Sets and functions.}
$\SetQ$, $\SetR$, $\SetR_+:=[0,\infty)$ are the sets of fractional,
real, and nonnegative real numbers, respectively. $\#\cal S$
denotes the number of elements in set $\cal S$, $\ln()$ the
natural and $\lb()$ the binary logarithm.

\noindent {\bf Asymptotics.}
We abbreviate $\lim_{n\to\infty}[f(n)-g(n)]=0$ by
$f(n)\toinfty{n}g(n)$ and say $f$ converges to $g$, without
implying that $\lim_{n\to\infty}g(n)$ itself exists. We write
$f(x)\leqm g(x)$ for $f(x)=O(g(x))$
and $f(x)\leqa g(x)$ for $f(x)\leq
g(x)+O(1)$.

\noindent {\bf Computability.}
A function $f:\S\to\SetR\cup\{\infty\}$ is said to be enumerable
(or lower semicomputable) if the set $\{(x,y)\,:\,y<f(x),\,
x\in\S,\, y\in\SetQ\}$ is recursively enumerable. $f$ is
co-enumerable (or upper semicomputable) if $[-f]$ is enumerable.
$f$ is computable (or estimable or recursive) if $f$ and $[-f]$ are
enumerable. $f$ is approximable (or limit-computable) if there is
a computable function $g:\S\times\SetN\to\SetR$ with
$\lim_{n\to\infty}g(x,n)=f(x)$.

\noindent {\bf Complexity.}
The conditional prefix (Kolmogorov) complexity
$K(x|y):=\min\{\l(p):U(y, p)=x \mbox{ halts} \}$ is the length of
the shortest binary program $p\in\B^*$ on a universal prefix
Turing machine $U$ with output $x\in\X^*$ and input $y\in\X^*$
\cite{Li:97}. $K(x):=K(x|\epstr)$.
For non-string objects $o$ we define
$K(o):=K(\langle o\rangle)$, where $\langle o\rangle\in\X^*$ is
some standard code for $o$. In particular, if $(f_i)_{i=1}^\infty$ is
an enumeration of all enumerable functions, we define
$K(f_i)=K(i)$.
We only need the following elementary properties: %
The co-enumerability of $K$, %
the upper bounds $K(x|\l(x))\leqa\l(x)\lb|\X|$ and $K(n)\leqa 2\lb n$, %
and $K(x|y)\leqa K(x)$, %
subadditivity $K(x)\leqa K(x,y)\leqa K(y)+K(x|y)$, and %
information non-increase $K(f(x))\leqa K(x)+K(f)$ for recursive $f:\X^*\to\X^*$. %

We need the concepts of (universal) (semi)measures for
strings \cite{Zvonkin:70}.

\fdefinition{defSemi}{(Semi)measures}{
We call $\nu:\X^*\to[0,1]$ a semimeasure if
$\nu(x)\geq\sum_{a\in\X}\nu(xa)\,\forall x\in\X^*$, and a
(probability) measure if equality holds and $\nu(\epstr)=1$.
$\nu(x)$ denotes the $\nu$-probability that a sequence starts with
string $x$. Further, $\nu(a|x):={\nu(xa)\over\nu(x)}$ is the
predictive $\nu$-probability that the next symbol is $a\in\X$,
given sequence $x\in\X^*$.
}

\fdefinition{defUniM}{Universal semimeasures $\MM$}{
A semimeasure $\MM$ is called a universal element of
a class of semimeasures $\M$, if it multiplicatively dominates
all members in the sense that
\beqn
  \mbox{ $\MM\in\M$ and
    $\forall\nu\in\M\;\exists w_\nu>0 : \MM(x)\geq w_\nu\!\cdot\!\nu(x)\;\forall x\in\X^*$.}
\eeqn
}
From now on we consider the (in a sense) largest class $\M$ which
is relevant from a constructive point of view (but see
\cite{Schmidhuber:00toe,Schmidhuber:02gtm,Hutter:03unipriors}
for even larger constructive classes), namely the class of {\em
all} semimeasures, which can be enumerated (=effectively be
approximated) from below:
\beq\label{Mclassdef}
  \M:= \;\mbox{class of all enumerable semimeasures}.
\eeq
Solomonoff \cite[Eq.(7)]{Solomonoff:64} defined the universal
predictor $\MM(y|x)=M(xy)/M(x)$ with $M(x)$ defined as the
probability that the output of a universal monotone Turing machine
starts with $x$ when provided with fair coin flips on the input
tape.
Levin \cite{Zvonkin:70} has shown that this $M$ is a universal
enumerable semimeasure.
%
Another possible definition of $M$ is as a (Bayes) mixture
\cite{Solomonoff:64,Zvonkin:70,Solomonoff:78,Li:97,Hutter:03unipriors,Hutter:04uaibook}:
$\tilde M(x)=\sum_{\nu\in\M}2^{-K(\nu)}\nu(x)$, where $K(\nu)$ is
the length of the shortest program computing function $\nu$.
Levin \cite{Zvonkin:70} has shown that the class of {\em all}
enumerable semimeasures is enumerable (with repetitions), hence
$\tilde M$ is enumerable, since $K$ is co-enumerable. Hence
$\tilde\MM\in\M$, which implies
\beq\label{Mdom}
  M(x) \;\geq\; w_{\tilde M} \tilde M(x)
  \;\geq\; w_{\tilde M} 2^{-K(\nu)}\nu(x)
  \;=\; w'_\nu\nu(x),
  \qmbox{where} w'_\nu\eqm 2^{-K(\nu)}.
\eeq
Up to a multiplicative constant, $\MM$ assigns higher probability
to all $x$ than any other enumerable semimeasure.
All $M$ have the same very slowly decreasing (in $\nu$)
domination constants $w'_\nu$, essentially because $\MM\in\M$.
We drop the prime from $w'_\nu$ in the following.
The mixture definition $\tilde M$ immediately generalizes to
arbitrary weighted sums of (semi)measures over countable
classes other than $\M$, but the class may not contain the mixture, and
the domination constants may be rapidly decreasing. We will
exploit this for the construction of the non-universal semimeasure
$W$ in Sections~\ref{secMLconv} and \ref{secNUESW}.

\section{Predictive Convergence with Probability 1}\label{secConv}

The following convergence results for $M$ are well-known
\cite{Solomonoff:78,Li:97,Hutter:02spupper,Hutter:04uaibook}.

\ftheorem{thConv}{Convergence of $M$ to $\mu$ w.p.1}{
For any universal semimeasure $M$ and any computable measure $\mu$ it holds:
\beqn
 \mbox{$M(x'_n|x_{<n}) \to \mu(x'_n|x_{<n})$
 for any $x'_n$ and
 ${M(x_n|x_{<n})\over\mu(x_n|x_{<n})} \to 1$, both
 w.p.1 for $n\to\infty$.}
\eeqn
}

The first convergence in difference is Solomonoff's
\cite{Solomonoff:78} celebrated convergence result. The second
convergence in ratio has first been derived by G\'acs
\cite{Li:97}.
Note the subtle difference between the two convergence results.
For {\em any} sequence $x'_{1:\infty}$ (possibly
constant and not necessarily random),
$M(x'_n|x_{<n})-\mu(x'_n|x_{<n})$ converges to zero w.p.1
(referring to $x_{1:\infty}$), but no statement is possible for
$M(x'_n|x_{<n})/\mu(x'_n|x_{<n})$, since
$\lim\,\inf\mu(x'_n|x_{<n})$ could be zero. On the other hand, if
we stay {\em on}-sequence ($x'_{1:\infty} = x_{1:\infty}$), we
have $M(x_n|x_{<n})/\mu(x_n|x_{<n}) \to 1$ (whether
$\inf\mu(x_n|x_{<n})$ tends to zero or not does not matter).
Indeed, it is easy to give an example where
$M(x'_n|x_{<n})/\mu(x'_n|x_{<n})$ diverges. For
$\mu(1|x_{<n})=1-\mu(0|x_{<n})=\odt n^{-3}$ we get
$\mu(0_{1:n})=\prod_{t=1}^n(1-\odt
t^{-3})\stackrel{n\to\infty}\longrightarrow c=0.450...>0$, i.e.\
$0_{1:\infty}$ is $\mu$-random. On the other hand, one can show
that $M(0_{<n})=O(1)$ and $M(0_{<n}1)\eqm 2^{-K(n)}$, which
implies ${M(1|0_{<n})\over \mu(1|0_{<n})} \eqm n^3\cdot
2^{-K(n)}\geqm n \to\infty$ for $n\to\infty$ ($K(n)\leqa
2\log n$).

Theorem~\ref{thConv} follows from (the discussion after) Lemma
\ref{lemHBounds} due to $M(x)\geq w_\mu\mu(x)$. Actually the Lemma
strengthens and generalizes Theorem~\ref{thConv}.
In the following we denote expectations w.r.t.\ measure $\rho$ by
$\E_\rho$, i.e.\ for a function $f:\X^n\to\SetR$,
$\E_\rho[f]=\sum'_{x_{1:n}}\rho(x_{1:n})f(x_{1:n})$, where $\sum'$
sums over all $x_{1:n}$ for which $\rho(x_{1:n})\neq 0$. Using
$\sum'$ instead $\sum$ is (only) important for partial functions $f$
undefined on a set of $\rho$-measure zero. Similarly $\P_{\!\rho}$
denotes the $\rho$-probability.

\flemma{lemHBounds}{Expected Bounds on Hellinger Sum}{
Let $\mu$ be a measure and $\nu$ be a semimeasure with $\nu(x)\geq
w\!\cdot\!\mu(x)$ $\forall x$. Then the following bounds on the
Hellinger distance $h_t(\nu,\mu|\o_{<t}) :=
\sum_{a\in\X}(\sqrt{\nu(a|\o_{<t})}-\sqrt{\mu(a|\o_{<t})}\,)^2$
hold:
\beqn
  \sum_{t=1}^\infty\E{\textstyle\left[\!\left(\sqrt{{\nu(\o_t|\o_{<t})
       \over\mu(\o_t|\o_{<t})}}\!-\!1\right)^2\right]}
  \;\stackrel{(i)}\leq\; \sum_{t=1}^\infty\E[h_t]
  \;\stackrel{(ii)}\leq\; 2\ln\{\E[\exp(\odt\sum_{t=1}^\infty h_t)]\}
  \;\stackrel{(iii)}\leq\; \ln w^{-1}
\eeqn
where $\E$ here and later means expectation w.r.t.\ $\mu$.
}

The $\ln w^{-1}$-bounds on the
first and second expression have first been derived in
\cite{Hutter:02spupper}, the second being a variation of
Solomonoff's bound
$\sum_n\E[(\nu(0|x_{<n})-\mu(0|x_{<n}))^2]\leq\odt\ln w^{-1}$.
If sequence $x_1x_2...$ is sampled from the probability measure
$\mu$, these bounds imply
\beqn
 \mbox{$\nu(x'_n|x_{<n}) \to \mu(x'_n|x_{<n})$
 for any $x'_n$ and
 ${\nu(x_n|x_{<n})\over\mu(x_n|x_{<n})} \to 1$, both
 w.p.1 for $n\to\infty$},
\eeqn
where w.p.1 stands here and in the following for `with
$\mu$-probability 1'.

Convergence is ``fast'' in the following sense: The second bound
($\sum_t\E[h_t]\leq \ln w^{-1}$) implies that the expected number
of times $t$ in which $h_t\geq\eps$ is finite and bounded by
${1\over\eps}\ln w^{-1}$. The new third bound represents a significant
improvement. It implies by means of a Markov inequality
that the probability of even only marginally exceeding this number
is extremely small, and that $\sum_t h_t$ is very unlikely to
exceed $\ln w^{-1}$ by much. More precisely:
\beqn\textstyle
  \P[\#\{t:h_t\geq\eps\}\geq{\textstyle{1\over\eps}}(\ln w^{-1}+c)]
  \;\leq\; \P[\sum_t h_t\geq \ln w^{-1}+c]
\eeqn\vspace{-3ex}
\beqn\textstyle
  \;=\; \P[\exp(\odt\sum_t h_t)\geq\e^{c/2}w^{-1/2}]
  \;\leq\; \sqrt{w}\E[\exp(\odt\sum_t h_t)]\e^{-c/2}
  \;\leq\; \e^{-c/2}.
\eeqn

\paradot{Proof}
We use the abbreviations $\rho_t=\rho(x_t|x_{<t})$
and $\rho_{1:n}=\rho_1\cdot...\cdot\rho_n=\rho(x_{1:n})$ for
$\rho\in\{\mu,\nu,R,N,...\}$ and
$h_t=\sum_{x_t}(\sqrt{\nu_t}-\sqrt{\mu_t})^2$.

$(i)$ follows from
\beqn
  \E[({\textstyle\sqrt{\nu_t\over\mu_t}}-1)^2|x_{<t}]
  \;\equiv \sum_{x_t:\mu_t\neq 0}\mu_t({\textstyle\sqrt{\nu_t\over\mu_t}}-1)^2
  = \sum_{x_t:\mu_t\neq 0}(\sqrt{\nu_t}-\sqrt{\mu_t})^2
  \;\leq\; h_t
\eeqn
by taking the expectation $\E[]$ and sum $\sum_{t=1}^\infty$.

$(ii)$ follows from Jensen's inequality
$\exp(\E[f])\leq\E[\exp{(f)}]$ for $f=\odt\sum_t h_t$.

$(iii)$ We exploit a construction
used in \cite[Thm.1]{Vovk:87}. For discrete (semi)measures
$p$ and $q$ with $\sum_i p_i=1$ and $\sum_i q_i\leq 1$ it holds:
\beq\label{eqsh}
  \sum_i\sqrt{p_i q_i}
  \;\leq\; 1-\odt\sum_i(\sqrt{p_i}-\sqrt{q_i})^2
  \;\leq\; \exp[-\odt\sum_i(\sqrt{p_i}-\sqrt{q_i})^2].
\eeq
The first inequality is obvious after multiplying out the second
expression. The second inequality follows from $1-x\leq\e^{-x}$.
Vovk \cite{Vovk:87} defined a measure $R_t:=\sqrt{\mu_t \nu_t}/N_t$
with normalization $N_t:=\sum_{x_t}\sqrt{\mu_t \nu_t}$.
Applying (\ref{eqsh}) for measure $\mu$ and semimeasure $\nu$ we get
$N_t\leq\exp(-\odt h_t)$. Together with $\nu(x)\geq w\cdot\mu(x)$
$\forall x$ this implies
\beqn
  \prod_{t=1}^n R_t
  \;=\; \prod_{t=1}^n {\sqrt{\mu_t \nu_t}\over N_t}
  \;=\; {\sqrt{\mu_{1:n}\nu_{1:n}}\over N_{1:n}}
  \;=\; \mu_{1:n} {\sqrt{\nu_{1:n}\over\mu_{1:n}}}N_{1:n}^{-1}
  \;\geq\; \mu_{1:n} \sqrt{w} \exp(\odt\sum_{t=1}^n h_t).
\eeqn
Summing over $x_{1:n}$ and exploiting $\sum_{x_t}R_t=1$ we get
$1\geq \sqrt{w}\E[\exp(\odt\sum_t h_t)$], which proves $(iii)$.

The bound and proof may be generalized to
$1\geq w^\kappa\E[\exp(\odt\sum_t\sum_{x_t}(\nu_t^\kappa-\mu_t^\kappa)^{1/\kappa})]$ with
$0\leq\kappa\leq\odt$ by defining
$R_t=\mu_t^{1-\kappa}\nu_t^\kappa/N_t$ with
$N_t=\sum_{x_t}\mu_t^{1-\kappa}\nu_t^\kappa$ and exploiting $\sum_i
p_i^{1-\kappa}q_i^\kappa \leq
\exp(-\odt\sum_i(p_i^\kappa-q_i^\kappa)^{1/\kappa})$.
\qed

One can show that the constant $\odt$ in Lemma~\ref{lemHBounds}
can essentially not be improved. Increasing it to a constant
$\alpha>1$ makes the expression infinite for some (Bernoulli)
distribution $\mu$ (however we choose $\nu$). For $\nu=M$ the
expression can become already infinite for $\alpha>\odt$ and some
computable measure $\mu$.

\section{Non-Convergence in Martin-L{\"o}f Sense}\label{secMLNconv}

Convergence of $M(x_n|x_{<n})$ to $\mu(x_n|x_{<n})$ with
$\mu$-probability 1 tells us that $M(x_n|x_{<n})$ is close to
$\mu(x_n|x_{<n})$ for sufficiently large $n$ on `most'
sequences $x_{1:\infty}$. It says nothing whether convergence is
true for any {\em particular} sequence (of measure 0).
Martin-L\"{o}f randomness can be used to capture convergence
properties for individual sequences.
Martin-L\"{o}f randomness is a very important and default concept of
randomness of individual sequences, which is closely related to
Kolmogorov complexity and Solomonoff's universal semimeasure
$M$. Levin gave a characterization equivalent to Martin-L\"{o}f's
original definition \cite{Levin:73random}:

\fdefinition{defML}{Martin-L\"{o}f random sequences}{
A sequence $\o=\o_{1:\infty}$ is $\mu$-Martin-L\"{o}f random
($\mu$.M.L.) iff there is a constant $c<\infty$ such that
$\MM(\o_{1:n})\leq c\cdot \mu(\o_{1:n})$ for all $n$.
Moreover, $d_\mu(\o):=\sup_n\{\lb {\MM(\o_{1:n})\over\mu(\o_{1:n})}\}\leq
\lb c$ is called the randomness deficiency of $\o$.
}
One can show that an M.L.-random sequence $x_{1:\infty}$ passes
{\em all} thinkable effective randomness tests, e.g.\ the law of
large numbers, the law of the iterated logarithm, etc. In
particular, the set of all $\mu$.M.L.-random sequences has
$\mu$-measure 1.

The open question we study in this section is whether $M$ converges
to $\mu$ (in difference or ratio) individually for all
Martin-L\"{o}f random sequences. Clearly, Theorem~\ref{thConv}
implies that convergence $\mu$.M.L. may at most fail for a set of
sequences with $\mu$-measure zero. A convergence M.L.\ result
would be particularly interesting and natural for $M$, since
M.L.-randomness can be defined in terms of $\MM$ itself (Definition
\ref{defML}).

The state of the art regarding this problem may be summarized as
follows: \cite{Vovk:87} contains a (non-improvable?) result which
is slightly too weak to imply M.L.-convergence,
\cite[Thm.5.2.2]{Li:97} and \cite[Thm.10]{Vitanyi:00} contain an
erroneous proof for M.L.-convergence, and
\cite{Hutter:03unipriors} proves a theorem indicating that the
answer may be hard and subtle (see \cite{Hutter:03unipriors} for
details).

The main contribution of this section is a partial answer to this
question. We show that M.L.-convergence fails at least for some
universal semimeasures:

\ftheorem{thMnonConv}{Universal semimeasure non-convergence}{
There exists a universal semimeasure $M$ and a computable measure
$\mu$ and a $\mu$.M.L.-random sequence $\alpha$, such that
\vspace{-2ex}\beqn
  M(\a_n|\a_{<n}) \;\;\not\!\!\!\longrightarrow \mu(\a_n|\a_{<n})
  \qmbox{for} n\to\infty.
\eeqn
}
This implies that also $M_n/\mu_n$ does not converge (since $\mu_n\leq
1$ is bounded). We do not know whether Theorem~\ref{thMnonConv}
holds for {\em all} universal semimeasures.
For the proof we need the concept of supermartingales. We only
define it for binary alphabet and uniform measure
$\mu(x)=\lambda(x):=2^{-\l(x)}$ for which we need it.

\fdefinition{defSemiMgale}{Supermartingale}{
\beqn
  \mbox{$m:\B^*\!\to\!\SetR$ is a supermartingale
  $:\Leftrightarrow$ $m(x)\geq\odt[m(x0)\!+\!m(x1)]$ for all $x\in\B^*$}
\eeqn
}
If $\nu$ is a (enumerable) semimeasure, then
$m:=\nu/\lambda$ is a (enumerable) supermartingale.
We prove the following theorem, which will imply Theorem
\ref{thMnonConv}.

\flemma{thSMnonConv}{Supermartingale non-convergence}{
For the M.L.-random sequence $\a$ defined in (\ref{defalpha}) and
the enumerable supermartingale $r$ defined in Lemma \ref{lemESM} and
for any $\eta,\eta'\in\SetR$ and any on $\alpha$ bounded
supermartingale $R$, i.e.\ $0<\eps<R(\alpha_{1:n})<c<\infty\,\forall
n$, it holds that
\beqn
  \left|{R(\a_{1:n})\over R(\a_{<n})}-\eta\right|>\delta
  \qmbox{or}
  \left|{R'(\a_{1:n})\over R'(\a_{<n})}-\eta'\right|>\delta
\eeqn
(or both) for a non-vanishing fraction of $n$, where
supermartingale $R':=\odt(R+r)$ and some $\delta>0$.
}

\paradot{Proof}
We define a sequence $\a$, which, in a sense, is the
lexicographically first (or equivalently left-most in the tree of
sequences) $\lambda$.M.L.-random sequence. Formally we define
$\a$, inductively in $n=1,2,3,...$ by
\beq\label{defalpha}
 \mbox{$\a_n=0$ if $M(\a_{<n}0)\leq 2^{-n}$, and $\a_n=1$ else.}
\eeq
We know that $M(\epstr)\leq 1$ and $M(\a_{<n}0)\leq 2^{-n}$ if
$\a_n=0$. Inductively, assuming $M(\a_{<n})\leq 2^{-n+1}$ for
$\a_n=1$ we have $2^{-n+1}\geq M(\a_{<n})\geq
M(\a_{<n}0)+M(\a_{<n}1) \geq 2^{-n}+M(\a_{<n}1)$ since $M$ is a
semimeasure, hence $M(\a_{<n}1)\leq 2^{-n}$.
Hence\footnote{Alternatively we may define $\a_n=0$ if
$M(0|\a_{<t})\leq\odt$ and $\a_n=1$ else.}
\beq\label{arand}
  \mbox{$M(\a_{1:n})\leq 2^{-n}\equiv\lambda(\a_{1:n})\,\forall n$,
  i.e.\ $\a$ is $\lambda$.M.L.-random.}
\eeq
With $R$ and $r$, also $R':=\odt(R+r)>0$ is a supermartingale. We
prove that the Theorem holds for infinitely many $n$. It is easy
to refine the proof to a non-vanishing fraction of $n$'s. Assume
that ${R(\a_{1:n})\over R(\a_{<n})}\to\eta$ for $n\to\infty$
(otherwise we are done). $\eta>1$ implies $R\to\infty$, $\eta<1$
implies $R\to 0$. Since $R$ is bounded, $\eta$ must be 1, hence
for sufficiently large $n_0$ we
have $|R(\a_{1:n})-R(\a_{<n})|<\eps$ for all $n\geq n_0$.

Assume $r\in\{0,\odt,1\}$ and $r(\a_{1:n})=\odt$ for infinitely
many $n$ and $r(\a_{1:n})=1$ for infinitely many $n$ (e.g.\ take
$r$ as defined in Lemma~\ref{lemESM}). Since $R$ stabilizes and $r$
oscillates, $R'$ cannot converge. Formally, for (the infinitely
many) $n\geq n_0$ for which $r(\a_{<n})=\odt$ and $r(\a_{1:n})=1$
we have
\beqn
  {R'(\a_{1:n})\over R'(\a_{<n})}-1
  \;\equiv\; {R(\a_{1:n})-R(\a_{<n})+r(\a_{1:n})-r(\a_{<n})\over R(\a_{<n})+r(\a_{<n})}
  \;\geq\; {-\eps+\odt\over c+\odt}
  \;\geq\; \delta \;>\; 0
\eeqn
for sufficiently small $\eps$ and $\delta$.
Similarly for (the infinitely many) $n\geq n_0$ for which
$r(\a_{<n})=1$ and $r(\a_{1:n})=\odt$ we have
\beqn
  1-{R'(\a_{1:n})\over R'(\a_{<n})}
  \;\equiv\; {R(\a_{<n})-R(\a_{1:n})+r(\a_{<n})-r(\a_{1:n})\over R(\a_{<n})+r(\a_{<n})}
  \;\geq\; {-\eps+\odt\over c+1}
  \;\geq\; \delta \;>\; 0.
\eeqn
This shows that Lemma~\ref{thSMnonConv} holds for infinitely
many $n$. If we define $r$ zero off-sequence, i.e.\ $r(x)=0$ for
$x\neq\a_{1:\l(x)}$, then $r$ is a supermartingale, but a
non-enumerable one, since $\alpha$ is not computable. In the next
lemma we define an enumerable supermartingale $r$, which completes
the proof of Lemma~\ref{thSMnonConv}. Finally note that we could
have defined $R'={R+\gamma r\over 1+\gamma}$ with arbitrarily
small $\gamma>0$, showing that already a small contamination can
destroy convergence. This is no longer true for the constructive
proof below.
\qed

\flemma{lemESM}{Enumerable supermartingale}{
Let $M^t$ with $t=1,2,3,...$ be computable approximations of $M$,
which enumerate $M$, i.e.\ $M^t(x)\nearrow M(x)$ for
$t\to\infty$. For each $t$ define recursively a sequence
$\alpha^t$ similarly to (\ref{defalpha}) as $\a^t_n=0$ if
$M^t(\a^t_{<n}0)\leq 2^{-n}$ and $\a^t_n=1$ else.
For even $\l(x)$ we define $r(x)=1$ if
$\exists t,n:x=\a^t_{<n}$ and $r(x)=0$ else.
For odd $\l(x)$ we define $r(x)=\odt[r(x0)+r(x1)]$.
$r$ is an enumerable supermartingale with
$r(\a_{1:n})$ being 1 and $\odt$ for a non-vanishing fraction of
$n$'s, where $\a=\lim_{t\to\infty}\a^t$ ($\a^t\!\!\nearrow\a$
lexicographically increasing).
}
The idea behind the definition of $r$ is to define $r(\a_{<n})=1$
for odd $n$ and if possible $\odt$ for even $n$.
The following possibilities exist for the local part of the sequence
tree:
\centerline{
$\mathop{\wedge}\limits^{r(x)}_{r(x0)\quad r(x1)}$
$=$
$\mathop{\wedge}\limits^0_{0\;\;\; 0}$,
$\l(x)$ odd
$\mathop{\wedge}\limits^{1/2}_{1\;\;\; 0}$ or
$\mathop{\wedge}\limits^{1/2}_{0\;\;\; 1}$ or
$\mathop{\wedge}\limits^1_{1\;\;\; 1}$, and
$\l(x)$ even
$\mathop{\wedge}\limits^1_{1/2\;\; 0}$ or
$\mathop{\wedge}\limits^1_{0\;\; 1/2}$ or
$\mathop{\wedge}\limits^1_{1/2\;\; 1/2}$,}
all
respecting the supermartingale property. The formal proof goes as follows:

\paradot{Proof}
$r$ is enumerable, since $\a^t_{<n}$ is computable. Further, $0\leq
r(x)\leq 1\,\forall x$. For odd $\l(x)$ the supermartingale property
$r(x)\geq\odt[r(x0)+r(x1)]$ is obviously satisfied. For even
$\l(x)$ and $x=\a^t_{<n}$ for some $t$ we have
$r(x)=1=\odt[1+1]\geq\odt[r(x0)+r(x1)]$. Even $\l(x)$ and
$x\neq\a^t_{<n}\,\forall t$ implies $xy\neq\a^t_{1:\l(xy)}\,\forall
t,y$, hence $r(x)=0=\odt[0+0]=\odt[r(x0)+r(x1)]$.
This shows that $r$ is a supermartingale.

Since $M^t$ is monotone increasing, $\a^t$ is also monotone
increasing w.r.t.\ to lexicographical ordering on $\B^\infty$.
Hence $\a^t_{1:n}$ converges to $\a_{1:n}$ for $t\to\infty$, and
even $\a^t_{1:n}=\a_{1:n}\,\forall t\geq t_n$ and sufficiently
large ($n$-dependent) $t_n$. This implies
$r(\a_{<n})=r(\a^{t_n}_{<n})=1$ for odd $n$. We know that $\a_n=0$
for a non-vanishing fraction of (even) $n$, since $\a$ is random.
For such $n$, $\a^t_n=0\,\forall t$, hence
$r(\a_{<n})=r(\a^{t_n}_{<n})=
\odt[r(\a^{t_n}_{<n}0)+r(\a^{t_n}_{<n}1)]=\odt[1+0]=\odt$. This shows that
$r(\a_{<n})=1$ ($\odt$) for a non-vanishing fraction of $n$,
namely the odd ones (the even ones with $\a_n=0$). \qed

\paradot{Nonconstructive Proof of Theorem~\ref{thMnonConv}}
Use Lemma~\ref{thSMnonConv} with $R:=M/\lambda$,
$R':=M'/\lambda$, $r=:q/\lambda$, hence $q$ is an enumerable
semimeasure, hence with $M$, also $M'=\odt(M+q)$ is a universal
semimeasure. $R(\a_{1:n})\leq 1$ from (\ref{arand}) and $R(x)\geq
c>0$ from universality of $M$ and computability of $\lambda$ show
that the conditions of Lemma~\ref{thSMnonConv} are satisfied.
Hence $R^(\!\,'\!\,^)(\a_{1:n})/R^(\!\,'\!\,^)(\a_{<n})\equiv
M^(\!\,'\!\,^)(\a_n|\a_{<n})/\lambda(\a_n|\a_{<n})\not\to 1$.
Multiplying this by $\lambda_n=\mu_n=\odt$ completes the proof. \qed

The proof of Theorem~\ref{thMnonConv} is non-constructive. Either
$M$ or $M'$ (or both) do not converge, but we do not know which
one. Below we give an alternative proof which is constructive. The
idea is to construct an enumerable (semi)measure $\nu$ such that
$\nu$ dominates $M$ on $\alpha$, but
$\nu(\a_n|\a_{<n})\not\to\odt$. Then we mix $M$ to $\nu$ to make
$\nu$ universal, but with larger contribution from $\nu$, in order
to preserve non-convergence.

\paradot{Constructive Proof of Theorem~\ref{thMnonConv}}
We define an enumerable semimeasure $\nu$ as follows:
\beq\label{nudef}
   \nu^t(x):=\left\{
   \begin{array}{ccl}
     2^{-t} & \mbox{if} & \l(x)=t \qmbox{and} x<\a_{1:t}^t \\
     0      & \mbox{if} & \l(x)=t \qmbox{and} x\geq\a_{1:t}^t \\
     0      & \mbox{if} & \l(x)>t \\
     \nu^t(x0)\!+\!\nu^t(x1) & \mbox{if} & \l(x)<t \\
   \end{array}\right.
\eeq
where $<$ is the lexicographical ordering on sequences, and
$\alpha^t$ has been defined in Lemma~\ref{lemESM}. $\nu^t$ is a
semimeasure, and with $\alpha^t$ also $\nu^t$ is computable and
monotone increasing in $t$, hence $\nu:=\lim_{t\to\infty}\nu^t$ is
an enumerable semimeasure (indeed, ${\nu(x)\over\nu(\epstr)}$ is a
measure). We could have defined a $\nu_{tn}$ by replacing
$\alpha_{1:t}^t$ with $\alpha_{1:t}^n$ in (\ref{nudef}). Since
$\nu_{tn}$ is monotone increasing in $t$ and $n$, any order of
$t,n\to\infty$ leads to $\nu$, so we have chosen arbitrarily $t=n$.
By induction (starting from $\l(x)=t$) it follows that
\beqn
  \nu^t(x)=2^{-\l(x)} \qmbox{if} x<\a_{1:\l(x)}^t \qmbox{and} \l(x)\leq t,
  \qquad\qquad
  \nu^t(x)=0 \qmbox{if} x>\a_{1:\l(x)}^t
\eeqn
On-sequence, i.e.\ for $x=\a_{1:n}$, $\nu^t$ is somewhere
in-between $0$ and $2^{-\l(x)}$. Since sequence $\a:=\lim_t\a^t$
is $\lambda$.M.L.-random it contains $01$ infinitely often,
actually $\a_n\a_{n+1}=01$ for a non-vanishing fraction of $n$. In
the following we fix such an $n$. For $t\geq n$ we get
\beqn
  \nu^t(\a_{<n})
  = \nu^t(\a_{<n}0) \!+\! \nu^t(\underbrace{\a_{<n}1}_{\nq\nq\nq>\a_{1:n}\geq\a_{1:n}^t,\text{ since }\a_n=0\nq\nq\nq})
  = \nu^t(\a_{<n}0)
  = \nu^t(\a_{1:n})
  \quad\Rightarrow\quad \nu(\a_{<n})=\nu(\a_{1:n})
\eeqn
This ensures $\nu(\a_n|\a_{<n})=1\neq\odt=\lambda_n$. For
$t>n$ large enough such that $\a_{1:n+1}^t=\a_{1:n+1}$ we get:
\beqn
  \nu^t(\a_{1:n})
  = \nu^t(\a_{1:n}^t)
  \geq \nu^t(\underbrace{\a_{1:n}^t 0}_{\nq\nq\nq<\a_{1:n+1}^t,\text{ since }\a_{n+1}=1\nq\nq\nq})
  = 2^{-n-1}
  \quad\Rightarrow\quad \nu(\a_{1:n})\geq 2^{-n-1}
\eeqn
This ensures $\nu(\a_{1:n})\geq 2^{-n-1}\geq\odt M(\a_{1:n})$ by
(\ref{arand}). Let $M$ be any universal semimeasure and
$0<\gamma<{1\over 5}$. Then $M'(x):=(1-\gamma)\nu(x)+\gamma
M(x)\,\forall x$ is also a universal semimeasure with
\bqan
  M'(\a_n|\a_{<n})
   & & \nq=\;\; {(1\!-\!\gamma)\nu(\a_{1:n})+\gamma M(\a_{1:n})\over (1\!-\!\gamma)\nu(\a_{<n})+\gamma M(\a_{<n})}
  \;\mathop{\rule{0ex}{2.5ex}\geq}^{\displaystyle\mathop{\rule{0ex}{2.5ex}
    \downarrow}^{\makebox[0ex]{\footnotesize
    $M(\a_{<n})\leq 2^{-n+1}$ and $M(\a_{1:n})\geq 0$}}}\;
  {(1\!-\!\gamma)\nu(\a_{1:n})\over (1\!-\!\gamma)\nu(\a_{<n})+\gamma 2^{-n+1}}
\\
   & & \nq\mathop=_{\displaystyle\mathop{\rule{0ex}{3ex}
    \uparrow}_{\rule{0ex}{2ex}\makebox[0ex]{\footnotesize
    $\nu(\a_{<n})=\nu(\a_{1:n})$}}}\;\;
  {1\!-\!\gamma\over 1\!-\!\gamma + \gamma 2^{-n+1}/\nu(\a_{1:n})}
  \;\;\mathop\geq_{\displaystyle\mathop{\rule{0ex}{3ex}
    \uparrow}_{\rule{0ex}{2ex}\makebox[0ex]{\footnotesize
    $\nu(\a_{1:n})\geq 2^{-n-1}$}}}\;\;
  {1\!-\!\gamma\over 1+3\gamma}
  \;\;>\;\; {1\over 2}.
\eqan
For instance for $\gamma={1\over 9}$ we have
$M'(\a_n|\a_{<n})\geq{2\over 3}\neq\odt=\lambda(\a_n|\a_{<n})$ for
a non-vanishing fraction of $n$'s. Note that the contamination of
$M$ with $\nu$ must be sufficiently large ($\gamma$ sufficiently
small), while an advantage of the the non-constructive proof
is that an arbitrarily small contamination sufficed.
\qed

A converse of Theorem~\ref{thMnonConv} can also be shown:

\ftheorem{thConvNR}{Convergence on nonrandom sequences}{
For every universal semimeasure $M$ there exist computable
measures $\mu$ and non-$\mu$.M.L.-random sequences $\a$ for which
$M(\a_n|\a_{<n})/\mu(\a_n|\a_{<n})\to 1$.
}

\section{Convergence in Martin-L{\"o}f Sense}\label{secMLconv}

In this section we give a positive answer to the question of
predictive M.L.-convergence to $\mu$. We consider general finite
alphabet $\X$.

\ftheorem{thMLconv}{Universal predictor for M.L.-random sequences}{
There exists an enumerable semimeasure $W$ such that for every
computable measure $\mu$ and every $\mu$.M.L.-random sequence $\omega$,
the predictions converge to each other:
\beqn\textstyle
  W(a|\o_{<t})\toinfty{t}\mu(a|\o_{<t}) \qmbox{for all} a\in\X \qmbox{if}
  d_\mu(\o)<\infty.
\eeqn
}

The semimeasure $W$ we will construct is not
universal in the sense of dominating all enumerable semimeasures,
unlike $M$. Normalizing $W$ shows that there is also a measure
whose predictions converge to $\mu$, but this measure is not
enumerable, only approximable. For proving Theorem~\ref{thMLconv}
we first define an intermediate measure $D$ as a mixture over all
computable measures, which is not even approximable. Based on
Lemmas
\ref{lemHBounds},\ref{lemHChain},\ref{lemE2I},
Proposition~\ref{proDtomu} shows that $D$ M.L.-converges to $\mu$.
We then define the concept of quasimeasures in Definition
\ref{defQM} and an enumerable semimeasure $W$ as a mixture over
all enumerable quasimeasures. Proposition~\ref{proWtoD} shows that
$W$ M.L.-converges to $D$. Theorem~\ref{thMLconv} immediately
follows from Propositions~\ref{proDtomu} and~\ref{proWtoD}.

\flemma{lemHChain}{Hellinger Chain}{
Let $h(\v p,\v q):=\sum_{i=1}^N(\sqrt{p_i}-\sqrt{q_i})^2$ be the
Hellinger distance between $\v p=(p_i)_{i=1}^N\in\SetR_+^N$ and
$\v q=(q_i)_{i=1}^N \in\SetR_+^N$. Then
\beqn
\begin{array}{rlcll}
  i)   & \mbox{for}\; \v p,\v q,\v r\in\SetR_+^N     & h(\v p,\v q)
       & \leq & (1+\beta)\,h(\v p,\v r)+(1+\beta^{-1})\,h(\v r,\v q), \;\mbox{any}\; \beta>0
\\
  ii)  & \mbox{for}\; \v p^1,...,\v p^m \in\SetR_+^N & h(\v p^1,\v p^m)
       & \leq & \displaystyle 3\sum_{k=2}^m k^2\,h(\v p^{k-1},\v p^k)
\end{array}
\eeqn
}

\paradot{Proof} $(i)$
For any $x,y,z\in\SetR$ and $\beta>0$, squaring the triangle
inequality $|x-y|\leq|x-z|+|z-y|$ and chaining it with the
binomial $2|x-z||z-y| \leq \beta(x-z)^2+\beta^{-1}(z-y)^2$ shows
$(x-y)^2\leq(1+\beta)(x-z)^2+(1+\beta^{-1})(z-y)^2$. $(i)$ follows
for $x=\sqrt{p_i}$, $y=\sqrt{q_i}$, and $z=\sqrt{r_i}$ and
summation over $i$.

$(ii)$ Applying $(i)$ for the triples $(\v p^k,\v p^{k+1},\v p^m)$
for and in order of $k=1,2,...,m-2$ with $\beta=\beta_k$ gives
\beqn
  h(\v p^1,\v p^m) \;\leq\;
  \sum_{k=2}^m \bigg[\prod_{j=1}^{k-2}(1\!+\!\beta_j^{-1})\bigg]
  \!\cdot\! (1\!+\!\beta_{k-1}) \!\cdot\! h(\v p^{k-1},\v p^k)
\eeqn
For $\beta_k=k(k+1)$ we have $\ln\prod_{j=1}^{k-2}(1+\beta_j^{-1}) \leq
\sum_{j=1}^\infty\ln(1+\beta_j^{-1}) \leq \sum_{j=1}^\infty \beta_j^{-1} = 1$
and $1+\beta_{k-1}\leq k^2$, which completes the proof. The choice
$\beta_k=2^{K(k)}$ would lead to a bound with $1+2^{K(k)}$ instead
of $k^2$. \qed

We need a way to convert expected bounds to bounds on individual
M.L.\ random sequences, sort of a converse of ``M.L.\ implies
w.p.1''. Consider for instance the Hellinger sum
$H(\o):=\sum_{t=1}^\infty h_t(\mu,\rho)/\ln w^{-1}$ between two
computable measures $\rho\geq w\!\cdot\!\mu$. Then $H$ is an
enumerable function and Lemma~\ref{lemHBounds} implies $\E[H]\leq
1$, hence $H$ is an integral $\mu$-test. $H$ can be increased to an enumerable
$\mu$-supermartingale $\bar H$. The universal $\mu$-supermartingale
$M/\mu$ multiplicatively dominates all enumerable supermartingales
(and hence $\bar H$).
Since $M/\mu\leq 2^{d_\mu(\o)}$, this implies the desired bound
$H(\o) \leqm 2^{d_\mu(\o)}$ for individual $\o$.
We give a self-contained direct proof, explicating all important
constants.

\flemma{lemE2I}{Expected to Individual Bound}{
Let $F(\o)\geq 0$ be an enumerable function and $\mu$ be an
enumerable measure and $\eps>0$ be co-enumerable. Then:
\beqn
  \qmbox{If} \E_\mu[F] \;\leq\; \eps
  \qmbox{then} F(\o) \;\leqm\; \eps\!\cdot\! 2^{K(\mu,F,\,^1\!/\eps)+d_\mu(\o)}
  \quad \forall\o
\eeqn
where $d_\mu(\o)$ is the $\mu$-randomness deficiency of $\o$ and
$K(\mu,F,\,^1\!/\eps)$ is the length of the shortest program for $\mu$,
$F$, and $^1\!/\eps$.
}

Lemma~\ref{lemE2I} roughly says that for $\mu$, $F$, and
$\eps\eqm\E_\mu[F]$ with short program ($K(\mu,F,^1\!/\eps)=O(1)$) and
$\mu$-random $\o$ ($d_\mu(\o)=O(1)$) we have $F(\o)\leqm
\E_\mu[F]$.

\paradot{Proof}
Let $F(\o)=\lim_{n\to\infty}F_n(\o)=\sup_n F_n(\o)$ be enumerated
by an increasing sequence of computable functions $F_n(\o)$.
$F_n(\o)$ can be chosen to depend on $\o_{1:n}$ only, i.e.\
$F_n(\o)=F_n(\o_{1:n})$ is independent of $\o_{n+1:\infty}$. Let
$\eps_n\!\!\searrow\eps$ co-enumerate $\eps$. We define
\beqn
  \bar\mu_n(\o_{1:k}) \;:=\; \eps_n^{-1} \nq\sum_{\o_{k+1:n}\in\X^{n-k}}\nq
  \mu(\o_{1:n})F_n(\o_{1:n}) \;\;\mbox{for}\;\; k\leq n,
  \qmbox{and} \bar\mu_n(\o_{1:k})=0 \;\;\mbox{for}\;\; k>n.
\eeqn
$\bar\mu_n$ is a computable semimeasure for each $n$ (due to
$\E_\mu[F_n]\leq\eps$) and increasing in $n$, since
\bqan
  \bar\mu_n(\o_{1:k}) & \!\!\geq\!\! & 0 \;=\; \bar\mu_{n-1}(\o_{1:k}) \qmbox{for} k\geq n \qmbox{and}
\\
  \bar\mu_n(\o_{<n}) & 
  \mathop\geq_{\displaystyle\mathop{\rule{0ex}{3.5ex}\nq
    \uparrow}_{\rule{0ex}{2ex}\makebox[0ex]{\footnotesize
      $\quad F_n\geq F_{n-1}$}}} &
  \!\!\sum_{\o_n\in\X} \eps_n^{-1}\mu(\o_{1:n})F_{n-1}(\o_{<n})
  \;\mathop=_{\displaystyle\mathop{\rule{0ex}{2.5ex}
    \uparrow}_{\rule{0ex}{2ex}\makebox[0ex]{\footnotesize $\mu$ measure}}}\;
  \eps_n^{-1}\mu(\o_{<n})F_{n-1}(\o_{<n})
  \;\mathop\geq_{\displaystyle\mathop{\rule{0ex}{2ex}
    \uparrow}_{\rule{0ex}{2ex}\makebox[0ex]{\footnotesize
      $\quad\eps_n\leq \eps_{n-1}$}}} \;
  \bar\mu_{n-1}(\o_{<n})
\eqan
and similarly for $k<n-1$. Hence $\bar\mu:=\bar\mu_\infty$ is an
enumerable semimeasure (indeed $\bar\mu$ is proportional to a
measure). From dominance (\ref{Mdom}) we get
\beq\label{eqMmuF}
  M(\o_{1:n})
  \;\geqm\; 2^{-K(\bar\mu)}\bar\mu(\o_{1:n})
  \;\geq\; 2^{-K(\bar\mu)}\bar\mu_n(\o_{1:n})
  \;=\; 2^{-K(\bar\mu)} \eps_n^{-1}\mu(\o_{1:n})F_n(\o_{1:n}).
\eeq
In order to enumerate $\bar\mu$, we need to enumerate $\mu$, $F$,
and $\eps^{-1}$, hence $K(\bar\mu)\leqa K(\mu,F,\,^1\!/\eps)$, so
we get
\beqn
  F_n(\o)\;\equiv\; F_n(\o_{1:n})
  \;\leqm\; \eps_n\!\cdot\! 2^{K(\mu,F,^1\!/\eps)}\!\cdot\!\textstyle{M(\o_{1:n})\over\mu(\o_{1:n})}
  \;\leq\; \eps_n\!\cdot\! 2^{K(\mu,F,^1\!/\eps)+d_\mu(\o)}.
\eeqn
Taking the limit $F_n\nearrow F$ and $\eps_n\!\!\searrow\eps$
completes the proof.
\qed

Let $\M=\{\nu_1,\nu_2,...\}$ be an enumeration of all enumerable
semimeasures, $J_k:=\{i\leq k \,:\, \nu_i $ is measure$\}$, and
$\delta_k(x):=\sum_{i\in J_k} \eps_i\nu_i(x)$. The weights
$\eps_i$ need to be computable and exponentially decreasing in $i$
and $\sum_{i=1}^\infty \eps_i\leq 1$. We choose $\eps_i=i^{-6}
2^{-i}$. Note the subtle and important fact that although the
definition of $J_k$ is non-constructive, as a finite set of finite
objects, $J_k$ is decidable (the program is unknowable for large
$k$). Hence, $\delta_k$ is computable, since enumerable measures
are computable.
\beqn
  D(x) \;=\; \delta_\infty(x)
  \;=\; \sum_{i\in J_\infty} \eps_i\nu_i(x)=
  \mbox{mixture of all computable measures}.
\eeqn
In contrast to $J_k$ and $\delta_k$, the set $J_\infty$ and hence
$D$ are neither enumerable nor co-enumerable. We also define the
measures $\hat\delta_k(x):=\delta_k(x)/\delta_k(\epstr)$ and $\hat
D(x):=D(x)/D(\epstr)$. The following Proposition implies predictive
convergence of $D$ to $\mu$ on $\mu$-random sequences.

\fproposition{proDtomu}{Convergence of incomputable measure $\hat D$}{
Let $\mu$ be a computable measure with index $k_0$, i.e.\
$\mu=\nu_{k_0}$. Then for the incomputable measure $\hat D$ and
the computable but non-constructive measures $\hat\delta_{k_0}$ defined above,
the following holds:
\beqn
\begin{array}{rccl}
  i)  & \sum_{t=1}^\infty h_t(\hat\delta_{k_0},\mu)    & \leqa & 2\ln 2 \!\cdot\! d_\mu(\o) +3k_0 \\[1ex]
  ii)   & \sum_{t=1}^\infty h_t(\hat\delta_{k_0},\hat D) & \leqm & k_0^7 2^{k_0+d_\mu(\o)} \\
\end{array}
\eeqn
}

Combining $(i)$ and $(ii)$, using Lemma~\ref{lemHChain}$(i)$, we
get $\sum_{t=1}^\infty h_t(\mu,\hat D) \leq c_\o f(k_0) < \infty$
for $\mu$-random $\o$, which implies $D(b|\o_{<t})\equiv\hat
D(b|\o_{<t}) \to \mu(b|\o_{<t})$. We do not know whether
on-sequence convergence of the ratio holds. Similar bounds hold
for $\hat\delta_{k_1}$ instead $\hat\delta_{k_0}$, $k_1\geq k_0$.
The principle proof idea is to convert the expected bounds of
Lemma~\ref{lemHBounds} to individual bounds, using Lemma
\ref{lemE2I}. The problem is that $\hat D$ is not computable,
which we circumvent by joining with Lemma~\ref{lemHChain}, bounds
on $\sum_th_t(\hat\delta_{k-1},\hat\delta_k)$ for
$k=k_0,k_0+1,...$.

\paradot{Proof}
$(i)$ Let $H(\o):=\sum_{t=1}^\infty h_t(\hat\delta_{k_0},\mu)$.
$\mu$ and $\hat\delta_{k_0}$ are measures with
$\hat\delta_{k_0}\geq \delta_{k_0}\geq \eps_{k_0}\mu$, since
$\delta_k(\epstr)\leq 1$, $\mu=\nu_{k_0}$ and $k_0\in J_{k_0}$.
Hence, Lemma~\ref{lemHBounds} applies and shows $\E_\mu[\exp(\odt
H)]\leq \eps_{k_0}^{-1/2}$.
$H$ is well-defined and enumerable for $d_\mu(\o)<\infty$, since
$d_\mu(\o)<\infty$ implies $\mu(\o_{1:t})\neq 0$ implies
$\hat\delta_{k_0}(\o_{1:t})\neq 0$. So $\mu(b|\o_{1:t})$ and
$\hat\delta_{k_0}(b|\o_{1:t})$ are well defined and computable
(given $J_{k_0}$). Hence $h_t(\hat\delta_{k_0},\mu)$ is
computable, hence $H(\o)$ is enumerable.
Lemma~\ref{lemE2I} then implies $\exp(\odt H(\o))\leqm
\eps_{k_0}^{-1/2}\cdot 2^{K(\mu,H,\sqrt{\eps}_{k_0})+d_\mu(\o)}$.
We bound
\beqn
  K(\mu,H,\sqrt{\eps}_{k_0})
  \;\leqa\; K(H|\mu,k_0) + K(k_0)
  \;\leqa\; K(J_{k_0}|k_0) + K(k_0)
  \;\leqa\; k_0 + 2\lb k_0.
\eeqn
The first inequality holds, since $k_0$ is the index and hence a
description of $\mu$, and $\eps_{()}$ is a simple computable
function. $H$ can be computed from $\mu$, $k_0$ and $J_{k_0}$,
which implies the second inequality. The last inequality follows
from $K(k_0)\leqa 2\lb k_0$ and the fact that for each $i\leq k_0$
one bit suffices to specify (non)membership to $J_{k_0}$, i.e.\
$K(J_{k_0}|k_0)\leqa k_0$. Putting everything together we get
\beqn
  H(\o) \;\leqa\; \ln\eps_{k_0}^{-1} + [k_0 + 2\lb k_0 + d_\mu(\o)]2\ln 2
  \;\leqa\; (2\ln 2) d_\mu(\o) + 3k_0.
\eeqn

$(ii)$ Let $H^k(\o):=\sum_{t=1}^\infty h_t(\hat\delta_k,\hat\delta_{k-1})$
and $k>k_0$. $\delta_{k-1}\leq\delta_k$ implies
\beqn
  {\hat\delta_{k-1}(x)\over\hat\delta_k(x)}
  \;\leq\; {\delta_k(\epstr)\over\delta_{k-1}(\epstr)}
  \;\leq\; {\delta_{k-1}(\epstr)+\eps_k\over\delta_{k-1}(\epstr)}
  \;=\; 1+{\eps_k\over\delta_{k-1}(\epstr)}
  \;\leq\; 1+{\eps_k\over\eps_O},
\eeqn
where $O:=\min\{i\in J_{k-1}\}=O(1)$. Note that $J_{k-1}\ni k_0$ is
not empty. Since $\hat\delta_{k-1}$ and $\hat\delta_k$ are
measures, Lemma~\ref{lemHBounds} applies and shows
$\E_{\hat\delta_{k-1}}[H^k]\leq
\ln(1+{\eps_k\over\eps_O})\leq {\eps_k\over\eps_O}$.
Exploiting $\eps_{k_0}\mu\leq\hat\delta_{k-1}$, this implies
$\E_\mu[H^k]\leq {\eps_k\over\eps_O\eps_{k_0}}$. Lemma
\ref{lemE2I} then implies $H^k(\o)\leqm
{\eps_k\over\eps_O\eps_{k_0}}\cdot
2^{K(\mu,H^k,\eps_O\eps_{k_0}/\eps_k)+d_\mu(\o)}$.
Similarly as in $(i)$ we can bound
\beqn
  K(\mu,H^k,\eps_{k_0}/\eps_O\eps_k) \leqa K(J_k|k)+K(k)+K(k_0) \leqa
  k+2\lb k+2\lb k_0, \qmbox{hence}
\eeqn\vspace{-3ex}
\beqn\textstyle
  H^k(\o) \;\leqm\; {\eps_k\over\eps_O\eps_{k_0}}\!\cdot\!k_0^2 k^2 2^k c_\o
  \;\eqm\; k_0^8 2^{k_0}k^{-4}c_\o, \qmbox{where} c_\o:=2^{d_\mu(\o)}.
\eeqn
Chaining this bound via Lemma~\ref{lemHChain}$(ii)$ we get for $k_1>k_0$:
\bqan
  \sum_{t=1}^n h_t(\hat\delta_{k_0},\hat\delta_{k_1})
   &\leq& \sum_{t=1}^n 3 \!\!\sum_{k=k_0+1}^{k_1}\!\! (k\!-\!k_0\!+\!1)^2 h_t(\hat\delta_{k-1},\hat\delta_k)
\\
  &\leq& 3 \!\!\sum_{k=k_0+1}^{k_1}\!\! k^2 H^k(\o)
  \;\leqm\; 3k_0^8 2^{k_0}c_\o \!\!\sum_{k=k_0+1}^{k_1}\!\! k^{-2}
  \;\leq\; 3k_0^7 2^{k_0}c_\o
\eqan
If we now take $k_1\to\infty$ we get $\sum_{t=1}^n
h_t(\hat\delta_{k_0},\hat D) \leqm 3k_0^7
2^{k_0+d_\mu(\o)}$. Finally let $n\to\infty$. \qed

The main properties allowing for proving $\hat D\to\mu$ were that
$\hat D$ is a measure with approximations $\hat\delta_k$, which
are computable in a certain sense. $\hat D$ is a mixture over all
enumerable/computable measures and hence incomputable.

\section{M.L.-Converging Enumerable Semimeasure $W$}\label{secNUESW}

The next step is to enlarge the class of computable measures to an
enumerable class of semimeasures, which are still sufficiently
close to measures in order not to spoil the convergence result.
For convergence w.p.1.\ we could include {\em all} semimeasures
(Theorem~\ref{thConv}). M.L.-convergence seems to require a more
restricted class. Included non-measures need to be zero on long
strings. We define quasimeasures as nearly normalized measures on $X^{\leq
n}$.

\fdefinition{defQM}{Quasimeasures}{
$\tilde\nu:\X^*\to\SetR_+$ is called a quasimeasure {\em iff}
$\tilde\nu$ is a measure or:
$\sum_{a\in\X}\tilde\nu(xa)=\tilde\nu(x)$ for $\l(x)<n$ and
$\tilde\nu(x)=0$ for $\l(x)>n$ and $1-{1\over
n}<\tilde\nu(\epstr)\leq 1$, for some $n\in\SetN$.
}

\flemma{lemQM}{Quasimeasures}{
$(i)$ A quasimeasure is either a semimeasure which is zero on long strings -or- a measure.
$(ii)$ The set of enumerable quasimeasures is enumerable and contains all
computable measures.
}

For enumerability it is important to include the measures in the
definition of quasimeasures. One way of enumeration would be
to enumerate all enumerable partial functions $f$ and convert them
to quasimeasures. Since we need a correspondence to
semimeasures, we convert a semimeasure $\nu$ directly
to a maximal quasimeasure $\tilde\nu\leq\nu$.

\paradot{Proof \& construction}
$(i)$ Obvious from Definition~\ref{defQM}.

$(ii)$ Let $\nu$ be an enumerable semimeasure enumerated by
$\nu^t\!\!\!\nearrow\!\nu$. Consider $m\equiv m^t := \max\{n\leq
t\,:\,\sum_{x_{1:n}}\nu^t(x_{1:n})>1-{1\over n}\}$. $m^t$ is finite
and monotone increasing in $t$. We define the
quasimeasure
\beqn
  \rho^t(x_{1:n}) := \sum_{x_{n+1:m}\in\X^{m-n}}\nu^t(x_{1:m})
  \qmbox{for} n\leq m \qmbox{and} \rho^t(x_{1:n})=0
  \qmbox{for} n>m.
\eeqn
We define an increasing sequence in $t$ of quasimeasures
$\tilde\nu^t\leq\nu^t$ for $t=1,2,...$ recursively
starting with $\tilde\nu^0:=0$ as follows:
\beqn
  \mbox{If $\rho^t(x_{1:n})\geq\tilde\nu^{t-1}(x_{1:n})$
$\forall x_{1:n}\forall n\leq m^t$ (and hence $\forall x$), then
$\tilde\nu^t:=\rho^t$, else $\tilde\nu^t:=\tilde\nu^{t-1}$.}
\eeqn
$\tilde\nu:=\lim_{t\to\infty}\tilde\nu^t$ is an enumerable
quasimeasure. Note that $m^\infty=\infty$ iff $\nu$ is a measure.
One can easily verify that $\tilde\nu\leq\nu$ and
$\tilde\nu\equiv\nu$ iff $\nu$ is a quasimeasure. This
implies that if $\nu_1,\nu_2,...$ is an enumeration of all
enumerable semimeasures, then $\tilde\nu_1,\tilde\nu_2,...$ is an
enumeration of all enumerable quasimeasures.
\qed

Let $\tilde\nu_1,\tilde\nu_2,...$ be the enumeration of all
enumerable quasimeasures constructed in the proof of Lemma
\ref{lemQM}, based on the enumeration of all enumerable
semimeasures $\nu_1,\nu_2,...$ with the property that
$\tilde\nu_i\leq\nu_i$ and equality holds if $\nu_i$ is a
(quasi)measure. We define the enumerable semimeasure
\beqn
  W(x):=\sum_{i=1}^\infty\eps_i\tilde\nu_i(x),
  \qmbox{and note that}
  D(x)=\sum_{i\in J} \eps_i\tilde\nu_i(x)
  \;\;\mbox{with}\;\;
  J:=\{i:\tilde\nu_i\mbox{ is measure}\}
\eeqn
with $\eps_i=i^{-6}2^{-i}$ as before.
To show $W\to D$ we need the following Lemma.

\flemma{lemHCont}{Hellinger Continuity}{
For $h_x(\mu,\nu):=\sum_{a\in\X}(\sqrt{\mu(a|x)}-\sqrt{\nu(a|x)})^2$,
where $\rho(y)=\mu(y)+\nu(y)$ $\forall y\in\X^*$ and $\mu$ and $\nu$
are semimeasures, it holds:
\beqn
\begin{array}{rl}
  i)   & h_x(\mu,\rho) \;\leq\; {\nu(x)\over\mu(x)}. \\[2mm]
  ii)  & h_x(\mu,\rho) \;\leq\; \odf\eps^2 \qmbox{if} \nu(x)\leq\eps\!\cdot\!\mu(x)
  \qmbox{and} \nu(xb)\leq\eps\!\cdot\!\mu(xb) \; \forall b\in\X. \\
\end{array}
\eeqn
}

$(ii)$ Since the Hellinger distance is locally quadratic, $h_x(\mu,\rho)$
scales quadratic in the deviation of predictor $\rho$ from $\mu$.
$(i)$ Closeness of $\rho(x)$ to $\mu(x)$ only, does not imply closeness
of the predicitons, hence only a bound linear in the deviation is possible.

\paradot{Proof}
$(i)$ We identify $\X\cong\{1,...,N\}$ and define $y_i=\mu(xi)$,
$z_i=\nu(xi)$, $y=\mu(x)$, and $z=\nu(x)$. We extend
$(y_i)_{i=1}^N$ to a probability by defining $y_0=y-\sum_{i=1}^N
y_i\geq 0$ and set $z_0=0$. Also $\eps':=z/y$.
Exploiting $\sum_{i=0}^N y_i = y$ and
$\sum_{i=0}^N z_i\leq z$ and $z\leq\eps y$ and $y_i,z_i,y,z\geq 0$
we get
\beqn
  h_x(\mu,\mu\!+\!\nu) \;\equiv\; \sum_{i=1}^N \Bigg(\sqrt{y_i\over y}-\sqrt{y_i\!+\!z_i\over y\!+\!z}\Bigg)^2
  \;\leq\; \sum_{i=0}^N \Bigg(\sqrt{y_i\over y}-\sqrt{y_i\!+\!z_i\over y\!+\!z}\Bigg)^2
\eeqn\vspace{-1ex}
\beqn
  \;=\; \sum_{i=0}^N \Bigg({y_i\over y}+{y_i\!+\!z_i\over y\!+\!z}-2\sqrt{y_i(y_i\!+\!z_i)\over y(y\!+\!z)}\Bigg)
  \;\leq\; 2-2\sum_{i=0}^N{y_i\over\sqrt{y(y\!+\!z)}}
  \;=\; 2-{2\over\sqrt{1\!+\!\eps'}}
  \;\leq\; \eps'.
\eeqn

$(ii)$ With the notation from $(i)$, additionally exploiting
$z_i\leq\eps y_i$ we get
\bqan
  \sqrt{y_i\!+\!z_i\over y\!+\!z}-\sqrt{y_i\over y}
  &\leq& {\sqrt{y_i\!+\!z_i}-\sqrt{y_i}\over\sqrt{y}}
  \;\leq\; {\sqrt{y_i(1\!+\!\eps)}-\sqrt{y_i}\over\sqrt{y}}
  \;\leq\; {\eps\over 2}\sqrt{y_i\over y}
  \qquad\mbox{and}
\\
  \sqrt{y_i\over y}-\sqrt{y_i\!+\!z_i\over y\!+\!z}
  &=& {\sqrt{y_i(1\!+\!\eps')}-\sqrt{y_i+z_i}\over\sqrt{y(1\!+\!\eps')}}
  \;\leq\; {\sqrt{y_i(1\!+\!\eps')}-\sqrt{y_i}\over\sqrt{y(1\!+\!\eps')}}
  \;\leq\; {\eps'\over 2}\sqrt{y_i\over y}.
\eqan
Exploiting $\eps'\leq\eps$, taking the square and summing over $i$ proves $(ii)$.
\qed

\fproposition{proWtoD}{Convergence of enumerable $W$ to incomputable $D$}{
For every computable measure $\mu$ and for $\o$ being
$\mu$-random, the following holds for $t\to\infty$:
\beqn
  (i)   \;\; {W(\o_{1:t})\over D(\o_{1:t})}\to 1, \qquad
  (ii)  \;\; {W(\o_t|\o_{<t})\over D(\o_t|\o_{<t})} \to 1, \qquad
  (iii) \;\; W(a|\o_{<t})\to D(a|\o_{<t}) \;\;\forall a\in\X.
\eeqn
}

The intuitive reason for the convergence is that the additional
contributions of non-measures to $W$ absent in $D$ are zero
for long sequences.

\paradot{Proof} $(i)$\vspace{-3ex}
\beq\label{eqWtoD}
  D(x) \;\leq\; W(x)
  \;=\; D(x) + \sum_{i\not\in J}\eps_i\tilde\nu_i(x)
  \;\leq\; D(x) + \sum_{i=k_x}^\infty\eps_i\tilde\nu_i(x),
\eeq
where $k_x:=\min_i\{i\not\in J:\tilde\nu_i(x)\neq 0\}$. For
$i\not\in J$, $\tilde\nu_i$ is not a measure. Hence
$\tilde\nu_i(x)=0$ for sufficiently long $x$. This implies
$k_x\to\infty$ for $\l(x)\to\infty$, hence $W(x)\to D(x)$ $\forall
x$. To get convergence in ratio we have to assume that
$x=\o_{1:n}$ with $\o$ being $\mu$-random, i.e.\
$c_\o:=\sup_n{M(\o_{1:n})\over\mu(\o_{1:n})}=2^{d_\mu(\o)}<\infty$.
\beqn
  \Rightarrow\; \tilde\nu_i(x)
  \;\leq\; \nu_i(x)
  \;\leq\; {1\over w_{\nu_i}}M(x)
  \;\leq\; {c_\omega\over w_{\nu_i}}\mu(x)
  \;\leq\; {c_\omega\over w_{\nu_i}\eps_{k_0}}D(x),
\eeqn
The last inequality holds, since $\mu$ is a computable measure of
index $k_0$, i.e.\ $\mu=\nu_{k_0}=\tilde\nu_{k_0}$. Inserting
$1/w_{\nu_i}\leq c'\cdot i^2$ for some $c=O(1)$ and
$\eps_i$ we get $\eps_i\tilde\nu_i(x) \leq {c'
c_\o\over\eps_{k_0}}i^{-4}2^{-i}D(x)$, which implies
$\sum_{i=k_x}^\infty \eps_i\tilde\nu_i(x) \leq \eps_x'D(x)$ with
\beqn
  \eps_x' \;:=\; {c'c_\o\over\eps_{k_0}}\sum_{i=k_x}^\infty i^{-4}2^{-i}
  \;\leq\; {2c'c_\o\over\eps_{k_0}}k_x^{-4}2^{-k_x} \to 0
  \qmbox{for} \l(x)\to\infty.
\eeqn
Inserting this into (\ref{eqWtoD}) we get
\beqn
  1 \;\leq\; {W(x)\over D(x)} \;\leq\; 1+\eps'_x \;\;\toinfty{\l(x)}\;\; 1
  \qmbox{for $\mu$-random $x$.}
\eeqn

$(ii)$ Obvious from $(i)$ by taking a double ratio.

$(iii)$
Since $D$ and $W-D$ are semimeasures and ${W-D\over W}\leq\eps'_x$
by $(i)$, Lemma~\ref{lemHCont}$(i)$ implies $h_x(D,W)\leq\eps'_x$.
Since $\eps'_x\to 0$ for $\mu$-random $x$, this shows $(iii)$.
$|W(a|x)-D(a|x)|\leq\eps'_x$ can also be shown.

\paradot{Speed of convergence}
The main convergence Theorem~\ref{thMLconv} now immediately
follows from Propositions~\ref{proDtomu} and~\ref{proWtoD}. We
briefly remark on the convergence rate.
For $M$, Lemma~\ref{lemHBounds} shows that $\E[\sum_t
h_t(M,\mu)]\leq\ln w_{k_0}^{-1}\eqm \ln k_0$ is logarithmic in the
index $k_0$ of $\mu$, but $\E[\sum_t
h_t(X,\mu)]\leq\ln\eps_{k_0}\eqm k_0$ is linear in $k_0$ for
$X=[W,D,\delta_{k_0}]$.
The individual bounds for $\sum_t h_t(\hat\delta_{k_0},\mu)$ and
$\sum_t h_t(\hat\delta_{k_0},\hat D)$ in Proposition
\ref{proDtomu} are linear and exponential in $k_0$, respectively.
For $W\stackrel{M.L.}\longrightarrow D$ we could not establish any
convergence speed.

Finally we show that $W$ does not dominate all enumerable
semimeasures, as the definition of $W$ suggests. We summarize all
computability, measure, and dominance properties of $M$, $D$,
$\hat D$, and $W$ in the following theorem:

\ftheorem{thMWDprop}{Properties of $M$, $W$, $D$, and $\hat D$}{\hspace{1ex} \\
$(i)$ $M$ is an enumerable semimeasure, which dominates all enumerable semimeasures. %
$M$ is not computable and not a measure. \\
$(ii)$ $\hat D$ is a measure, $D$ is proportional to a measure, both
dominating all enumerable quasimeasures. %
$D$ and $\hat D$ are not computable %
and do not dominate all enumerable semimeasures. \\
$(iii)$ $W$ is an enumerable semimeasure, which dominates all enumerable quasimeasures. %
$W$ is not itself a quasimeasure, is not computable, and does not dominate all enumerable semimeasures.
}

We conjecture that $D$ and $\hat D$ are not even approximable
(limit-computable), but lie somewhere higher in the arithmetic
hierarchy. Since $W$ can be normalized to an approximable measure
M.L.-converging to $\mu$, and $D$ was only an intermediate
quantity, the question of approximability of $D$ seems not
too interesting.

\paradot{Proof}
$(i)$ First sentence: Holds by definition. That such an $M$
exists follows from the enumerability of all enumerable
semimeasures \cite{Zvonkin:70,Li:97}.
Second sentence: If $M$ were a measure it would be computable,
contradicting \cite[Thm.4$(iii)$]{Hutter:03unipriors} (see below).

$(ii)$ First sentence: Follows from the definition of $D$ and
$\hat D$ and the fact that quasimeasures are zero on long strings:
${D\over\nu}\geq\eps_\nu>0$ if $\nu$ is a computable measure. If
$\nu$ is a ``proper'' quasimeasure, then
$\min_{x\in\X^*}{D(x)\over\nu(x)}=\min_{x:\l(x)\leq
m_\nu}{D(x)\over\nu(x)}>0$, since $\nu(x)=0$ for
$\l(x)>m_\nu<\infty$, and $D(x)>0\,\forall x$.
Second sentence:
It is well known that there is no computable semimeasure
dominating all computable measures (see e.g.\
\cite[Thm.4]{Hutter:03unipriors}), which shows that $D$, $\hat
D$ and $W$ cannot be computable.
We now show that $D$ and $W$ do not dominate the enumerable
semimeasure $M$ by extending this argument. Let $\nu$ be a
nowhere\footnote{$M$, $W$, $\hat D$, $D$, and $\delta_k$ for
$k\geq O(1)$ are nowhere zero. Alternatively one can verify that
all relevant assertions remain valid if $\nu$ is somewhere zero.}
zero computable semimeasure.
We define a computable sequence $\a$ as follows by induction:
Given $\a_{<n}$, choose some $\a_n$ in a computable way (by
computing $\nu$ to sufficient accuracy) such that
$\nu(\a_n|\a_{<n})<|\X|^{-1}(1+{1\over n^2})$.
Such an $\a_n$ exists, since $\nu$ is a semimeasure.
We then define the computable deterministic measure $\bar\nu$
concentrated on $\a$, i.e.\ $\bar\nu(\a_{1:n})=1$ $\forall n$ and
$\bar\nu(x)=0$ for all $x$ which are not prefixes of $\a$. By the
chain rule we get $\nu(\a_{1:n})\leq
{\sinh\,\pi\over\pi}|\X|^{-n}\leq 4 |\X|^{-n}\bar\nu(\a_{1:n})$.
This shows that no computable semimeasure $\nu$ can dominate all
computable measures, since $\bar\nu$ is not dominated. We use this
construction for $\nu=\delta_k$:
\beqn
  \sum_{i=1}^k \eps_i\tilde\nu_i(\a_{1:n})
  \;\;\mathop{\rule{0ex}{2.5ex}=}^{\displaystyle\mathop{\rule{0ex}{2.5ex}
    \downarrow}^{\makebox[0ex]{\footnotesize
    for sufficiently large $n=n_k$}}}\;\;
  \; \delta_k(\a_{1:n})
  \;\leq\; 4 |\X|^{-n}\bar\delta_k(\a_{1:n})
  \;\;\mathop{\rule{0ex}{2.5ex}\leqm}^{\displaystyle\mathop{\rule{0ex}{2ex}
    \downarrow}^{\makebox[0ex]{\footnotesize
    $M\geqm 2^{-K(\nu)}\nu$}}}\;\;
 |\X|^{-n} 2^{K(\bar\delta_k)}M(\a_{1:n})
\eeqn\vspace{-2ex}
\beq\label{eqDNUni}
   \;\;\mathop{\leqm}_{\displaystyle\mathop{\rule{0ex}{2ex}
    \uparrow}_{\rule{0ex}{2ex}\makebox[0ex]{\footnotesize
    $K(\bar\delta_k)\leqa K(\delta_k)\leqa k+2\lb k$}}}\;\;
  |\X|^{-n} k^2 2^k M(\a_{1:n})
  \;\;\mathop\leq_{\displaystyle\mathop{\rule{0ex}{3ex}
      \uparrow}_{\rule{0ex}{2ex}\makebox[0ex]{\footnotesize
      for $n\geq{2\over\lb|\X|}k$}}}\;\;
  k^2 2^{-k} M(\a_{1:n}).
\eeq
For all $x$ we have
\beqn
  D(x)-\delta_k(x) \;\leq\; \!\sum_{i=k+1}^\infty\! \eps_i\tilde\nu_i(x)
  \;=\; \!\sum_{i=k+1}^\infty\! i^{-6}2^{-i} \tilde\nu_i(x)
  \;\leq\; 2^{-k} \!\sum_{i=k+1}^\infty\! i^{-6} \nu_i(x)
  \;\leqm\; 2^{-k}M(x).
\eeqn
Summing both bounds we get
$
  D(\a_{1:n_k}) \leq W(\a_{1:n_k})
  \leqm (k^2+1)2^{-k} M(\a_{1:n_k}) $, which shows that $D$, $\hat D$
and $W$ do not dominate the enumerable semimeasure $M$.

Remark: Note that the constructed sequence(s) $\a$ depends on the
choice of $k$, so we should write more precisely $\a=\a^k$. For
$D$ (but not for $W$) we can choose $k={n\over 2}\lb|\X|$ in
(\ref{eqDNUni}) (satisfying $n\geq{2\over\lb|\X|}k$), leading to
$D(\a^n_{1:n})\leqm n^2|\X|^{-n/2}M(\a^n_{1:n})$. It is easy to
generalize (\ref{eqDNUni}) to $\forall
x_{<t}\exists\a_{t:n}:\delta_k(x_{<t}\a_{t:n})\leqm|\X|^{t-n}k^2
2^k M(x_{<t}\a_{t:n})$, where $t$ is a simple function of $k$.
Choosing $t=k^2+1$ and $n=(k+1)^2$ and joining the results for
$k=1,2,...$ and $x_{<t}:=\a_{<t}$ we get $D(\a_{1:n})\leqm n
2^{-\sqrt n} M(\a_{1:n})\,\forall n$ for the single sequence $\a$.
This implies that (but is stronger than) $\a$ is not random
w.r.t.\ to any computable measure $\tilde\nu$. Such $\a$ are
sometimes called absolutely non-stochastic.

$(iii)$ First sentence: Enumerability is immediate from the
definition, given the enumerability of all enumerable
quasimeasures.
Second sentence: Since quasimeasures drop out in the mixture
defining $W$ for long $x$, $W$ cannot be a measure. Since
$W(x)\neq 0\,\forall x$ it is also not a quasimeasure.
Non-computability and non-dominance of $W$ have already been shown
in $(ii)$. \qed

\section{Conclusions}\label{secDisc}

We investigated a natural strengthening of Solomonoff's famous
convergence theorem, the latter stating that with probability 1
(w.p.1) the prediction of a universal semimeasure $M$ converges to
the true computable distribution $\mu$
($M\stackrel{w.p.1}\longrightarrow\mu$).
We answered partially negative the question of whether convergence
also holds individually for all Martin-L{\"o}f (M.L.) random
sequences
($\exists M : M${\scriptsize\tabcolsep=2pt
\begin{tabular}{c} \\[-12pt] $M.L.$ \\[-3pt] $\not\!\!\longrightarrow$
\end{tabular}}$\mu$).
We constructed random sequences $\a$ for which there exist
universal semimeasures on which convergence fails.
Multiplicative dominance of $M$ is the key property to show
convergence w.p.1. Dominance over all measures is also satisfied
by the restricted mixture $W$ over all quasimeasures. We showed
that $W$ converges to $\mu$ on all M.L.-random sequences by
exploiting the incomputable mixture $D$ over all measures. For
$D\stackrel{M.L.}\longrightarrow\mu$ we achieved a (weak)
convergence rate; for $W\stackrel{M.L.}\longrightarrow D$ and
$W/D\stackrel{M.L.}\longrightarrow 1$ only an asymptotic result.
The convergence rate properties w.p.1.\ of $D$ and $W$ are as
excellent as for $M$.

We do not know whether $D/\mu\stackrel{M.L.}\longrightarrow 1$
holds. We also do not know the convergence rate for
$W\stackrel{M.L.}\longrightarrow D$, and the current bound for
$D\stackrel{M.L.}\longrightarrow\mu$ is double exponentially worse
than for $M\stackrel{w.p.1}\longrightarrow\mu$. A minor question
is whether $D$ is approximable (which is unlikely).
Finally there could still exist {\em universal} semimeasures $M$
(dominating all enumerable semimeasures) for which
M.L.-convergence holds ($\exists M :
M\stackrel{M.L.}\longrightarrow\mu\,?$). In case they exist, we
expect them to have particularly interesting additional structure
and properties.
While most results in algorithmic information theory are
independent of the choice of the underlying universal Turing
machine (UTM) or universal semimeasure (USM), there are also
results which depend on this choice. For instance, one can show
that $\{(x,n):K_U(x)\leq n\}$ is tt-complete for some $U$, but
not tt-complete for others \cite{Muchnik:02}. A
potential $U$ dependence also occurs for predictions based on
monotone complexity \cite{Hutter:03unimdl}.
It could lead to interesting insights to identify a class of
``natural'' UTMs/USMs which have a variety of favorable
properties. A more moderate approach may be to consider classes
${\cal C}_i$ of UTMs/USMs satisfying certain properties ${\cal
P}_i$ and showing that the intersection $\cap_i {\cal C}_i$ is not
empty.

Another interesting and potentially fruitful approach to the
convergence problem at hand is to consider other classes of
semimeasures $\M$, define mixtures $M$ over $\M$, and (possibly)
generalized randomness concepts by using this $M$ in Definition
\ref{defML}. Using this approach, in \cite{Hutter:03unipriors} it
has been shown that convergence holds for a subclass of Bernoulli
distributions if the class is dense, but fails if the class is
gappy, showing that a denseness characterization of $\M$ could be
promising in general.

\paradot{Acknowledgements}
We want to thank Alexey Chernov for his invaluable help.


\begin{small}

\end{small}


\begin{thebibliography}{Hut03d}

\bibitem[HM04]{Hutter:04mlconvx}
M.~Hutter and An.~A. Muchnik.
\newblock Universal convergence of semimeasures on individual random sequences.
\newblock In {\em Proc. 15th International Conf. on Algorithmic Learning Theory
  ({ALT'04})}, volume 3244 of {\em LNAI}, pages 234--248, Padova, 2004.
  Springer, Berlin.

\bibitem[Hut03a]{Hutter:02spupper}
M.~Hutter.
\newblock Convergence and loss bounds for {Bayesian} sequence prediction.
\newblock {\em IEEE Transactions on Information Theory}, 49(8):2061--2067,
  2003.

\bibitem[Hut03b]{Hutter:03unipriors}
M.~Hutter.
\newblock On the existence and convergence of computable universal priors.
\newblock In {\em Proc. 14th International Conf. on Algorithmic Learning Theory
  ({ALT'03})}, volume 2842 of {\em LNAI}, pages 298--312, Sapporo, 2003.
  Springer, Berlin.

\bibitem[Hut03c]{Hutter:03mlconv}
M.~Hutter.
\newblock An open problem regarding the convergence of universal a priori
  probability.
\newblock In {\em Proc. 16th Annual Conf. on Learning Theory ({COLT'03})},
  volume 2777 of {\em LNAI}, pages 738--740, Washington, DC, 2003. Springer,
  Berlin.

\bibitem[Hut03d]{Hutter:03unimdl}
M.~Hutter.
\newblock Sequence prediction based on monotone complexity.
\newblock In {\em Proc. 16th Annual Conf. on Learning Theory ({COLT'03})},
  volume 2777 of {\em LNAI}, pages 506--521, Washington, DC, 2003. Springer,
  Berlin.

\bibitem[Hut05]{Hutter:04uaibook}
M.~Hutter.
\newblock {\em Universal Artificial Intelligence: Sequential Decisions based on
  Algorithmic Probability}.
\newblock Springer, Berlin, 2005.
\newblock 300 pages, http://www.idsia.ch/$_{^{\sim}}$marcus/ai/uaibook.htm.

\bibitem[Lev73]{Levin:73random}
L.~A. Levin.
\newblock On the notion of a random sequence.
\newblock {\em Soviet Mathematics Doklady}, 14(5):1413--1416, 1973.

\bibitem[LV97]{Li:97}
M.~Li and P.~M.~B. Vit\'anyi.
\newblock {\em An Introduction to {K}olmogorov Complexity and its
  Applications}.
\newblock Springer, Berlin, 2nd edition, 1997.

\bibitem[ML66]{MartinLoef:66}
P.~Martin-L{\"o}f.
\newblock The definition of random sequences.
\newblock {\em Information and Control}, 9(6):602--619, 1966.

\bibitem[MP02]{Muchnik:02}
An.~A. Muchnik and S.~Y. Positselsky.
\newblock {Kolmogorov} entropy in the context of computability theory.
\newblock {\em Theoretical Computer Science}, 271(1--2):15--35, 2002.

\bibitem[Sch00]{Schmidhuber:00toe}
J.~Schmidhuber.
\newblock Algorithmic theories of everything.
\newblock Report IDSIA-20-00, quant-ph/0011122, {IDSIA}, Manno (Lugano),
  Switzerland, 2000.

\bibitem[Sch02]{Schmidhuber:02gtm}
J.~Schmidhuber.
\newblock Hierarchies of generalized {Kolmogorov} complexities and
  nonenumerable universal measures computable in the limit.
\newblock {\em International Journal of Foundations of Computer Science},
  13(4):587--612, 2002.

\bibitem[Sol64]{Solomonoff:64}
R.~J. Solomonoff.
\newblock A formal theory of inductive inference: Parts 1 and 2.
\newblock {\em Information and Control}, 7:1--22 and 224--254, 1964.

\bibitem[Sol78]{Solomonoff:78}
R.~J. Solomonoff.
\newblock Complexity-based induction systems: Comparisons and convergence
  theorems.
\newblock {\em IEEE Transactions on Information Theory}, IT-24:422--432, 1978.

\bibitem[VL00]{Vitanyi:00}
P.~M.~B. Vit\'anyi and M.~Li.
\newblock Minimum description length induction, {B}ayesianism, and {K}olmogorov
  complexity.
\newblock {\em IEEE Transactions on Information Theory}, 46(2):446--464, 2000.

\bibitem[Vov87]{Vovk:87}
V.~G. Vovk.
\newblock On a randomness criterion.
\newblock {\em Soviet Mathematics Doklady}, 35(3):656--660, 1987.

\bibitem[ZL70]{Zvonkin:70}
A.~K. Zvonkin and L.~A. Levin.
\newblock The complexity of finite objects and the development of the concepts
  of information and randomness by means of the theory of algorithms.
\newblock {\em Russian Mathematical Surveys}, 25(6):83--124, 1970.

\end{thebibliography}
\end{document}